\documentclass[sort&compress]{elsarticle}

\usepackage{amssymb,amsmath,amsthm}
\usepackage{color}
\usepackage{bm}
\usepackage[hidelinks]{hyperref}

\usepackage{graphicx}
\graphicspath{{pics/}}

\usepackage{mmacells}
\mmaSet{
  morefv={gobble=1},
  linklocaluri=mma/symbol/definition:#1,
  leftmargin=\parindent+21pt
}

\usepackage{feynmp-auto}

\usepackage[capitalize]{cleveref}

\usepackage{booktabs,tabularx}

\usepackage{fullpage}

\newcommand{\SummerTime}{\texttt{SummerTime}}
\newcommand{\DREAM}{\texttt{DREAM}}
\newcommand{\Mathematica}{\textit{Mathematica}}
\newcommand{\WMathematica}{\textit{Wolfram Mathematica}}
\newcommand{\LiteRed}{\texttt{LiteRed}}

\newcommand{\TFunction}{\texttt{TFunction}}
\newcommand{\TRecurrence}{\texttt{TRecurrence}}

\newcommand{\TMasters}{\texttt{TMasters}}

\newcommand{\D}{d}
\newcommand{\Sum}{\mathop{\Sigma}\nolimits}
\newcommand{\SumDown}{\Sum_{<} }
\newcommand{\SumUp}{\Sum_{\ge} }

\renewcommand{\Im}{\operatorname{Im}}

\begin{document}

\begin{frontmatter}

\title{\DREAM, a program for arbitrary-precision computation of dimensional recurrence relations solutions, and its applications}

\author[binp]{Roman N. Lee}
\ead{r.n.lee@inp.nsk.su}
\author[binp]{Kirill T. Mingulov}
\ead{k.t.mingulov@inp.nsk.su}

\address[binp]{Budker Institute of Nuclear Physics, 630090, Novosibirsk, Russia}

\begin{abstract}
We present the \Mathematica\ package \DREAM\ for arbitrarily high precision computation of multiloop integrals within the DRA (Dimensional Recurrence \& Analyticity) method as solutions of dimensional recurrence relations.
Starting from these relations, the package automatically constructs the inhomogeneous solutions and reduces the manual efforts to setting proper homogeneous solutions.
\DREAM\ also provides means to define the homogeneous solutions of the higher-order recurrence relations (and can construct those of the first-order recurrence relations automatically).
Therefore, this package can be used to apply the DRA method to the topologies with sectors having more than one master integral.
Two nontrivial examples are presented:
four-loop fully massive tadpole diagrams of cat-eye topology and
three-loop cut diagrams which are necessary for computation of the width of the para-positronium decay into four photons.
The analytical form of this width is obtained here for the first time to the best of our knowledge.
\end{abstract}


\end{frontmatter}

\begin{small}
\noindent
{\em Manuscript Title:} DREAM, a program for arbitrary-precision computation of dimensional recurrence relations solutions, and its applications                                      \\
{\em Authors:} Roman N. Lee, Kirill T. Mingulov               \\
{\em Program Title:} DREAM                               \\
{\em Journal Reference:}                                      \\
{\em Catalogue identifier:}                                   \\
{\em Licensing provisions:} none                               \\
{\em Programming language:} Wolfram Mathematica                \\
{\em Computer:} Any with Wolfram Mathematica installed\\
{\em Operating system:} Any supporting Wolfram Mathematica     \\
{\em RAM:} Depending on the complexity of the problem         \\
{\em Number of processors used:} Depending on the problem (one or all available) 
{\em Classification:}  4.4 Feynman diagrams, 4.7 Other Functions, 5 Computer Algebra.                                        \\
{\em Nature of problem:}\\
Construction and evaluation of the multiloop integrals as solutions of difference equations systems by space-time dimensionality $d$. \\
{\em Solution method:}\\
Inhomogeneous solutions are automatically built from the matrix of the difference equations; homogeneous solutions are fixed within the DRA method and provided by the user.
Arising multiple nested sums have a factorized summand and can be computed in polynomial time.
For harmonic sums, the convergence acceleration technique is applied. 
The precision of the results obtained by the package is sufficient for the application of the integer relation algorithm \texttt{PSLQ}. \\
{\em Restrictions:} Depending on the complexity of the problem, limited by RAM and CPU time. \\
{\em Running time:} From several seconds to several hours, depending on the complexity of the problem.\\
\end{small}

\section*{Introduction}

The contemporary collider experiments give rise to interest in multiloop corrections and in multiloop computations methods in general.
Analytical multiloop computation methods use so-called Integration-by-Parts identities (IBP identities) \cite{ChetTka1981,Tkachov1981}, which allow one to reduce a Feynman diagram to a finite set of simpler diagrams, which are often referred to as master integrals.

In the case when the integrals non-trivially depend on some parameter (which can be, e.g., the ratio of two kinematic invariants), one can use the differential equations approach \cite{Kotikov1991b,Remiddi1997}.
If there is no such a parameter, this method cannot help directly.
One can use instead the difference equations for master integrals.
Both the differential and difference equations can be derived using the IBP reduction.
The method formulated in Ref.~\cite{Laporta2000} is based on the difference equations with respect to the power of one of the massive propagators.
Recently it was used in Ref.~\cite{Laporta2017} for the high-precision computation of the 4-loop QED contributions to anomalous magnetic moment of the electron.
In Ref.~\cite{Tarasov1996}, it was proposed to use dimensional recurrence relations.
In Ref.~\cite{Lee2010}, the method of calculation based on these relations and analytical properties of multiloop integrals as functions of space-time dimensionality was introduced (the DRA method).

In this method, the multiloop integrals are considered as functions of $\D$ and the difference equations connecting integrals in $\D$ and $\D+2$ dimensions are derived from the Feynman parametrization.
Using the analytical properties of integrals with respect to $\D$, it is possible to fix the arbitrary periodic functions in the general solution up to several constants, which should be calculated by means of other methods.
The solutions are given in terms of nested sums with the factorized summand.
The factorized form of the summand in the sums permits their efficient numerical computation.
In particular, the dedicated \Mathematica\ package \SummerTime\ has been published recently \cite{LeeMingulov:2016:SummerTime}.
This program allows one to compute ``scalar'' sums, which arise from difference equations systems with triangular matrix in the right-hand side.
However, as the complexity of the integral families grows, one also encounters the systems with non-triangular matrices.
Two obstacles appear when one tries to apply the DRA method to such systems.
First, one should be able to construct the meromorphic homogeneous solutions of the higher-order recurrence relations.
In contrast to the case of the first-order recurrence relations, this problem is highly non-trivial.
We present an approach to the solution of this problem in a parallel paper \cite{LeeMingulov:Meromorphic}.
The second obstacle is high-precision numerical calculation of the sums which summand is factorized only in the sense of matrix multiplication.

In the present paper we propose the \WMathematica\ package \DREAM\ (which stands for Dimensional REcurrence \& Analyticity Method).
This package is dedicated to application of the DRA method to the difference equations systems with non-triangular matrix in the right-hand side (of course, it can also treat the triangular case).
This package can deal with nested sums which appear in this case and automates the construction of the inhomogeneous parts of the solutions.
It can find automatically the homogeneous solutions of the first-order recurrence relations with the prescribed analytical properties (in terms of $\Gamma$-functions).
It also contains tools for the custom definitions of the homogeneous solutions of the higher-order recurrence relations.
In addition to the extension of the applicability region of the DRA method, we also had in mind the goal of alleviating the access to this method for the wide community involved in multiloop calculations.


The paper is organized as follows.
In the first section we review the DRA method.
In the next section we describe the main ideas lying in the foundation of the \DREAM\ package.
The third section contains examples of the package use.
The package is supplemented with two examples involving non-triangular matrices of recurrence relations:
the master integrals entering the width of parapositronium 4$\gamma$-decay and four-loop fully massive tadpole master integrals (some results for this family are presented in the appendix).
More applications will surely come in time;
for instance, \DREAM\ appears to be applicable to the computation of four-loop form-factors \cite{HennSmirnovSteinhauser:4loop-form-factors,HennLeeSmirnovSteinhauser:4loop-form-factors,LeeSmirnovSteinhauser:4loop-form-factors-nf2,ManteuffelSchabinger:4loop-form-factors-nf3}.
The fourth section provides a brief review of \DREAM's functionality.
In the conclusion, the main results of the present paper are given.

\section{Dimensional recurrence relations}

The dimensional recurrence relations for master integrals can be obtained using IBP-identities and Feynman parametrization.
They have the form:
\begin{equation}
	\mathbf{J}(\nu+1) = \mathbb{M}(\nu) \, \mathbf{J}(\nu),
\end{equation}
where $\nu=\D/2$, $\mathbf{J} = (J_1, \ldots, J_N)$ is a column-vector of master integrals,
and $\mathbb{M}$ is a square matrix with coefficients being rational functions of $\nu$.
Assuming that $\mathbb{M}$ has the block-triangular form:
\begin{equation}
	\mathbb{M} = \begin{pmatrix}
		\mathbb{M}_{11} & 0 & \ldots & 0 \\
		\mathbb{M}_{21} & \mathbb{M}_{22} & \ldots & 0 \\
		\vdots & \vdots & \ddots & \vdots \\
		\mathbb{M}_{K1} & \mathbb{M}_{K2} & \ldots & \mathbb{M}_{KK}
	\end{pmatrix},
\end{equation}
we can rewrite the equations for the integrals of $k$-th block as:
\begin{equation}
	\label{eq:drr-system}
	\mathbf{J}_k(\nu+1) - \mathbb{M}_{kk}(\nu) \, \mathbf{J}_k(\nu) = \mathbf{R}_k(\nu) = \sum_{i=1}^{k-1} \mathbb{M}_{ki}(\nu) \, \mathbf{J}_i(\nu).
\end{equation}
Each diagonal block $\mathbb{M}_{kk}$ in $\mathbb{M}$ corresponds to master integrals from the specific sector.
The right-hand side of this equation contains the contributions of master integrals from subsectors (i.e.\ the sectors obtained by contraction of some lines).
The structure of the difference equations can be visualized as an oriented graph with block indices being vertices;
vertices $k$ and $i$ are connected by the oriented edge $k \to i$ if the corresponding matrix block $\mathbb{M}_{ki}$ is non-zero.
We will refer to this graph as \emph{dependency graph} $G$.
The graph $G$ has a natural interpretation in terms of sectors: the edge from $k$ to $i$ in $G$ is equivalent to the statement that the $i$-th sector is a subsector of the $k$-th sector.

Now we will turn to construction of the 	solution of the difference equation system~\eqref{eq:drr-system}.
Let $\mathbb{H}_k(\nu)$ be a homogeneous solution of the $k$-th matrix block:
\begin{equation}
	\label{eq:homo-part}
	\mathbb{H}_k(\nu+1) = \mathbb{M}_{kk}(\nu) \, \mathbb{H}_k(\nu).
\end{equation}
Note that the homogeneous solution can be trivially found from \cref{eq:homo-part} for the scalar case (when $\mathbb{M}_{kk}$ is a $1\times1$ matrix).
In the general matrix case, there exists no common algorithm for homogeneous solution construction.
In the parallel paper \cite{LeeMingulov:Meromorphic}, we present a method to accomplish this task.
In what follows, we will not consider this problem and will assume that the homogeneous solutions are already known.

The inhomogeneous solution for \cref{eq:drr-system} can be expressed as
\begin{equation}
	\mathbf{J}_k(\nu) = \mathbb{H}_{k}(\nu) \Sum \mathbb{H}_k^{-1}(\nu+1) \mathbf{R}_k(\nu),
\end{equation}
where $\Sum$ denotes the operator of indefinite summation, so that
\begin{equation}
	(\Sum f)(\nu+1) - (\Sum f)(\nu) = f(\nu).
\end{equation}
This operator can be defined as
\begin{equation}
\label{eq:sum-up}
\SumUp f(\nu) = -\sum_{n=0}^\infty f(\nu+n)
\end{equation}
or
\begin{equation}
\label{eq:sum-down}
\SumDown f(\nu) = \sum_{n=-\infty}^{-1} f(\nu+n),
\end{equation}
depending on which sum converges.
If the integrals from the right-hand side are found within the same framework,
the final result can be represented as
\begin{equation}
	\label{eq:drr-sol}
	\mathbf{J}_k(\nu) = \mathbb{H}_k(\nu) \, \bm{\Omega}_k(\nu) + \sum_{\bm{i} \in G} \mathbb{S}_{\bm{i}}(\nu) \, \bm{\Omega}_{i_0}(\nu),
\end{equation}
where
\begin{multline}
\label{eq:fold-sum}
\mathbb{S}_{\bm{i}} = \mathbb{H}_{i_{N}} \Sum_{\tau_{N}} \mathbb{H}_{i_{N}}^{-1} \,
\cdots \,
\Sum_{\tau_{2}} \mathbb{H}_{i_2}^{-1} \mathbb{T}_{i_2 i_1} \mathbb{H}_{i_1} \,
\Sum_{\tau_{1}} \mathbb{H}_{i_1}^{-1} \mathbb{T}_{i_1 i_0} \mathbb{H}_{i_0} \\
= (-1)^{\#(\geq)} \, \mathbb{H}_{i_{N}}(\nu) \sum_{k_{N} \tau_{N} 0} 
\mathbb{H}_{i_{N}}^{-1}(\nu+k_{N}) \, \mathbb{T}_{i_{N}, i_{N-1}}(\nu+k_{N}) \cdots
\mathbb{H}_{i_2}(\nu+k_{3}) \sum_{k_{2} \tau_{2} k_{3}}
\mathbb{H}_{i_2}^{-1}(\nu+k_{2}) \, \mathbb{T}_{i_2, i_1}(\nu+k_2) \\
\times \mathbb{H}_{i_1}(\nu+k_{2}) \sum_{k_{1} \tau_{1} k_{2}}
\mathbb{H}_{i_1}^{-1}(\nu+k_{1}) \mathbb{T}_{i_1, \, i_0}(\nu+k_{1})
\mathbb{H}_{i_0}(\nu+k_{1})
\end{multline}
and the sum in \cref{eq:drr-sol} goes over all paths $\bm{i}=i_N \to i_{N-1} \to \ldots \to i_0$ in $G$ starting from the vertex $i_N=k$.
The symbols $\tau_j \in \{ \geq, < \}$ determine the summation direction and $\#(\geq)$ is a number of the symbols $\geq$ in the list $(\tau_1, \ldots, \tau_N)$.
The rectangular matrices $\mathbb{T}_{ij}$ (referred to as ``adapters'' below) are expressed in terms of matrix blocks of $\mathbb{M}$ as
\begin{equation}
	\label{eq:adapter}
	\mathbb{T}_{ij}(\nu) = \mathbb{M}^{-1}_{ii}(\nu) \, \mathbb{M}_{ij}(\nu).
\end{equation}
The entries of $\bm{\Omega}_j$ are arbitrary periodic functions for the $j$-th block which should be fixed within the DRA method.

The standard treatment of the solution of \cref{eq:drr-sol} is the high-precision numerical calculation of the coefficients of the $\epsilon$-expansion followed by the application of \texttt{PSLQ} algorithm \cite{FergusonBailey:1992:PSLQ}.
If all matrices in \cref{eq:fold-sum} are scalars (i.e.\ $1\times1$ matrices), this task of numerical computation can be effectively solved by \SummerTime\ package.
However, if nontrivial matrix blocks appear, the generalization of \SummerTime\ algorithms is necessary, and this was the main motivation for the development of the \DREAM\ package.
The approach used in \SummerTime\ is to construct the oriented tree graph related to the sum \eqref{eq:fold-sum}.
In particular, when $\tau_1=\ldots=\tau_{N}$, this graph is nothing else but $\bm{i} = i_N \to i_{N-1} \to \ldots \to i_0$.
Then the sums related to the nodes of this tree are calculated starting from the leaves.
Another natural approach is to represent nodes of this tree as objects which communicate with each other by the messages traveling along the edges of the graph (in both directions).
In addition, it allows one to create custom objects which are needed to define homogeneous solutions for nontrivial blocks.
It is this approach that is chosen in the \DREAM\ package.

\section{Objects and messages in \DREAM}

In \DREAM, each node in the oriented tree corresponding to the sum \eqref{eq:fold-sum} is represented as an object which is associated with the sum of the following form:
\begin{equation}
	\label{eq:fold-sum-representation}
	\mathbb{S}(\nu) = \mathbb{H}(\nu)
		\Sum_{\tau} \mathbb{H}^{-1}(\nu) \, \mathbb{T}(\nu) \, \widetilde{\mathbb{S}}(\nu),
\end{equation}
where $\tau \in \{ \geq, < \}$ and $\Sum_{\geq,<}$ are defined in \cref{eq:sum-up,eq:sum-down},
$\mathbb{H}$ is a homogeneous solution, \cref{eq:homo-part}, and $\mathbb{T}$ is an adapter, \cref{eq:adapter}.
The quantity $\widetilde{\mathbb{S}}$ can be either of the same form \eqref{eq:fold-sum-representation} or some homogeneous solution $\widetilde{\mathbb{H}}$.
This object has a few methods for the computation of the corresponding sum and some properties containing the essential information about the summand (in particular, its decay rate).

The object is ready to compute the associated sum only after its child nodes have computed theirs.
Thus, the process of computation can be organized via the messages which go along the edges of the tree.

If one aims at calculating the sum \eqref{eq:fold-sum} with some precision $p$, the object representing the outer sum by $i_N=k$ receives the corresponding request.
Then it determines the working precision $p_w$ and number of terms $n_w$ which are	 required for the calculation.
In order to determine these parameters, it is necessary to analyze the asymptotics of the summand (both \DREAM\ and \SummerTime\ have empirical formulae for their estimation).
Then the object sends request messages to its children to compute the summand with precision $p_w$ at $n_w$ successive points.
In turn, these objects determine their own working parameters $p'_w$ and $n'_w$ and send $(p'_w, n'_w)$ to their own children.
In such a manner, the tree will be traversed from its root to its leaves and working parameters will be known for all computational algorithms of all objects.
Another traversal of the tree allows one to calculate the sum \eqref{eq:fold-sum}:
starting from the leaves of the tree, objects compute their associated sums and send the messages with the results to their parents.
In turn, the parents use these results for the computation of their own associated sums and send messages to their own parents, and so on.

Let us now determine the overall number of the objects which are to be created and computed.
Since each sum $\mathbb{S}_{\bm{i}}$ is associated with a tree, the naive estimate for the number of objects would be $\sum_{\bm{i}\in G} \#(\text{nodes in }\bm{i})$.
However, it is clear that some objects belonging to different sums can be reused.
In terms of dependencies, it is natural to have one object per one node of $G$,
however since the working parameters $p_w$ and $n_w$ depend on the continuation of the path (via the asymptotics of the summand), it appears to be necessary to have one object per one path $\bm{i}$ in $G$.

\section{Examples of applications}

In this section we present two examples of \DREAM\ applications.

\subsection{Master integrals for the width of positronium decay in 4$\gamma$}

In this example we compute the master integrals entering the width of para-positronium decay into four photons at the leading order by the fine structure constant $\alpha$.
It has been known from theoretical calculations made by several groups \cite{Billoire:1978:pPs4gamma-LO,MutaNiuya:1982:pPs4gamma-LO,Lepage:1983:pPs4gamma-LO,AdkinsPfahl:1999:pPs4gamma-NLO}
and so far the most precise computation has been made by Adkins and Pfahl~\cite{AdkinsPfahl:1999:pPs4gamma-NLO}:
\begin{equation}
	\label{eq:adkins-pfahl}
	\Gamma^{\text{(LO)}}(\mathrm{Ps}\to4\gamma) = \frac{m\alpha^5}{2} \left(\frac{\alpha}{\pi}\right)^2 0.274290(8).
\end{equation}
The precision of this result is, of course, sufficient for any practical purposes, but not sufficient to recover the analytical form of $\Gamma^{\text{(LO)}}(\mathrm{Ps}\to4\gamma)$.
The \DREAM\ package allows us to compute the master integrals entering this quantity with high precision and then apply the \texttt{PSLQ} algorithm \cite{FergusonBailey:1992:PSLQ} in order to obtain the expression for $\Gamma^{\text{(LO)}}(\mathrm{Ps}\to4\gamma)$ in terms of conventional transcendental numbers.

\begin{figure}
	\centering
	\includegraphics[scale=0.5]{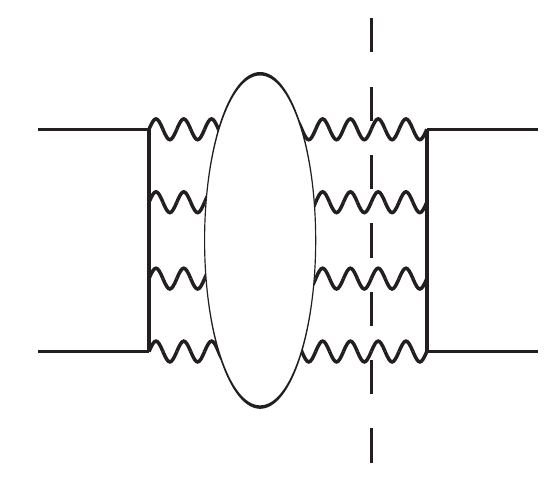}
	\caption{$\Gamma^{\text{(LO)}}(\mathrm{Ps} \to 4\gamma)$ diagram topology. The ellipsis denotes all possible permutations of the photon lines.}
	\label{pic:ps-topology}
\end{figure}

\begin{figure}
	\centering
	\setlength{\tabcolsep}{0pt}
	\renewcommand{\arraystretch}{0.5}
	\newcommand{\miscale}{0.4}
	\newcommand{\misize}{\footnotesize}
	\begin{tabular}{ccccc}
		\includegraphics[scale=\miscale]{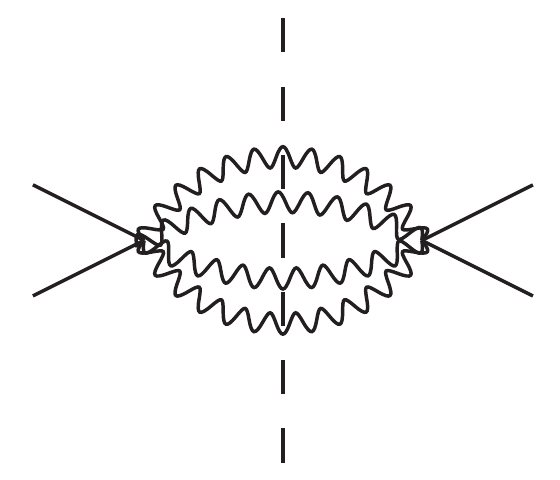} &
		\includegraphics[scale=\miscale]{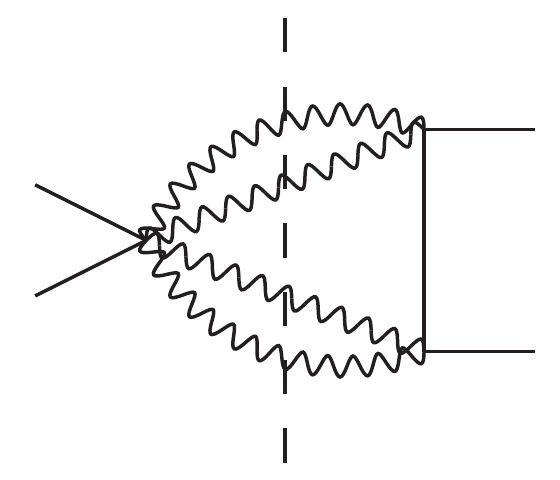} &
		\includegraphics[scale=\miscale]{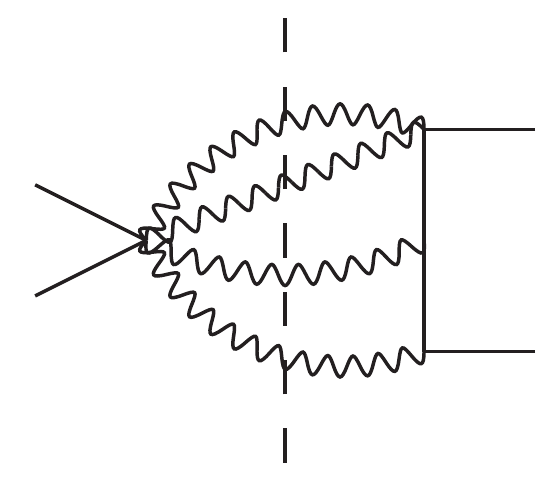} &
		\includegraphics[scale=\miscale]{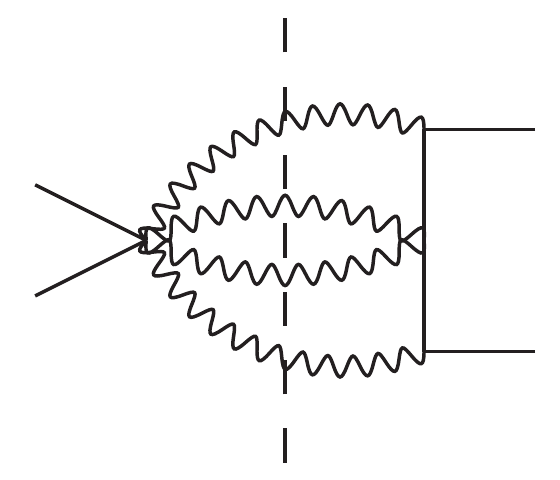} &
		\includegraphics[scale=\miscale]{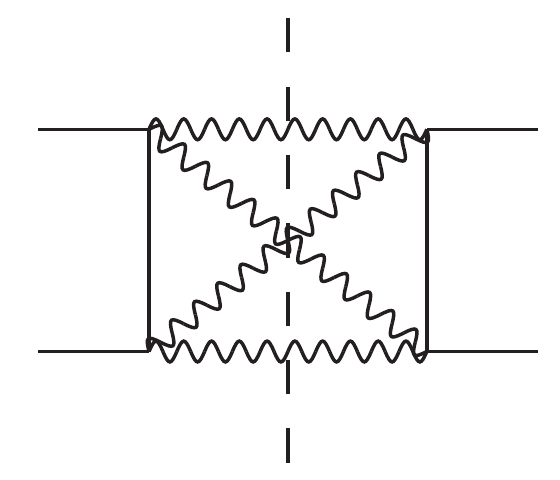} \\
		\misize $\mathrm{Ps}_1$ &
		\misize $\mathrm{Ps}_2$ &
		\misize $\mathrm{Ps}_3$ &
		\misize $\mathrm{Ps}_4$ &
		\misize $\mathrm{Ps}_5$ \\
		\hspace{-2pt}\includegraphics[scale=\miscale]{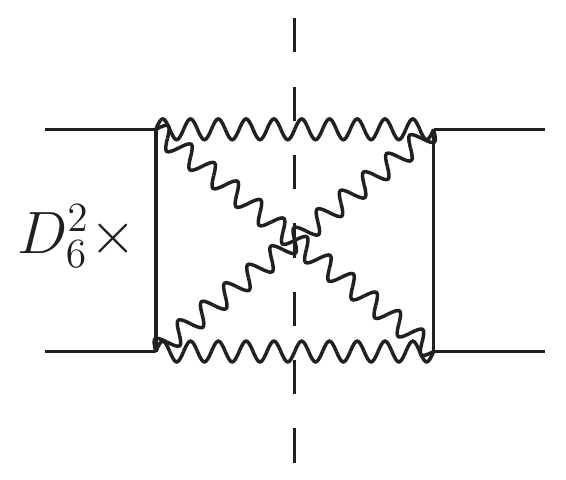} &
		\includegraphics[scale=\miscale]{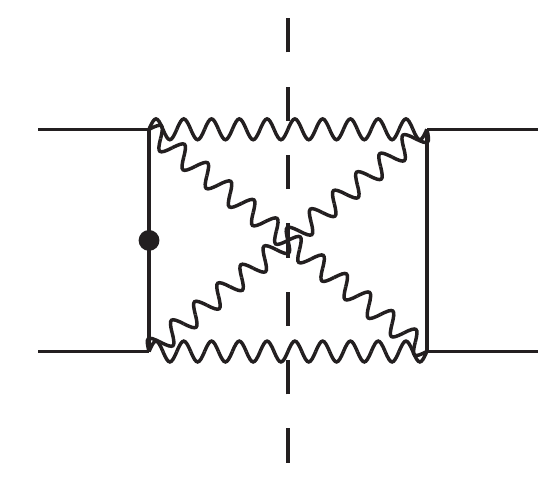} &
		\includegraphics[scale=\miscale]{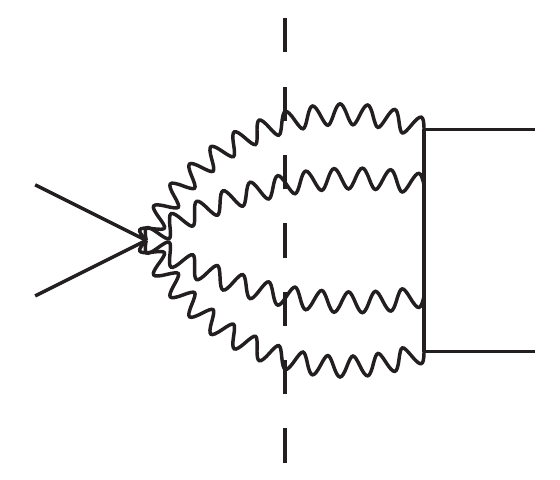} &
		\includegraphics[scale=\miscale]{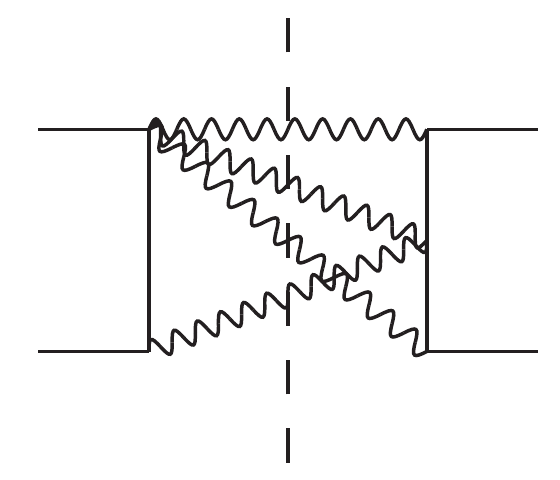} &
		\includegraphics[scale=\miscale]{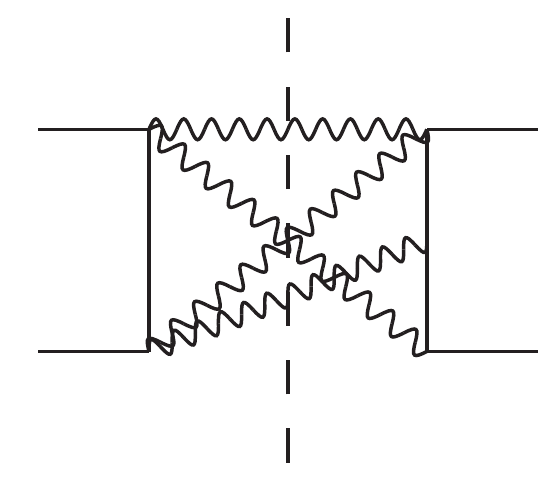} \\
		\misize $\mathrm{Ps}_6$ &
		\misize $\mathrm{Ps}_7$ &
		\misize $\mathrm{Ps}_8$ &
		\misize $\mathrm{Ps}_9$ &
		\misize $\mathrm{Ps}_{10}$
	\end{tabular}
	\setlength{\tabcolsep}{6pt}
	\renewcommand{\arraystretch}{1.0}
	\caption{
		Master-integrals $\mathrm{Ps}_{1, \ldots, 10}$.
		Dot on a line stands for a squared propagator.
	}
	\label{pic:ps-masters}
\end{figure}

For the computation of the decay width, one can consider diagrams of form as in Fig.~\ref{pic:ps-topology} with four photon lines cut.
These diagrams can be expressed in terms of scalar integrals of the following form:
\begin{equation}
	\label{eq:ps-topology}
	\mathrm{Ps}(n_1, \ldots, n_9) =
		\int \frac{\mathrm{d}l \, \mathrm{d}k \, \mathrm{d}q}{\pi^{3d/2}} \,
		\theta(l^0, k^0, q^0, l^0+k^0+q^0-2p^0)
		\prod_{i=1}^4 \Im\frac{1}{(D_i + i0)^{n_i}} \prod_{i=5}^9 \frac{1}{(D_i + i0)^{n_i}},
\end{equation}
where $D_i$ are the following denominators:
\begin{equation}
\begin{aligned}
	D_{1} &= l^2, & 
	D_{2} &= k^2, &
	D_{3} &= q^2, \\ 
	D_{4} &= (l+k+q-2p)^2, &
	D_{5} &= (l+k-p)^2-1, &
	D_{6} &= (l+k+q-p)^2-1, \\
	D_{7} &= (q-p)^2-1, &
	D_{8} &= (l+q-p)^2-1, &
	D_{9} &= (k-p)^2-1. \\
\end{aligned}
\end{equation}%
Using the \LiteRed\ package \cite{Lee:2013mka}, we reduced the diagram topology to the set of the ten master integrals depicted on Fig.~\ref{pic:ps-masters}:
\begin{equation}
	\begin{aligned}
		\mathrm{Ps}_1    &= \mathrm{Ps}(1, 1, 1, 1, 0, 0, 0, 0, 0), &
		\mathrm{Ps}_2    &= \mathrm{Ps}(1, 1, 1, 1, 0, 0, 0, 1, 0), \\
		\mathrm{Ps}_3    &= \mathrm{Ps}(1, 1, 1, 1, 0, 0, 0, 1, 1), &
		\mathrm{Ps}_4    &= \mathrm{Ps}(1, 1, 1, 1, 0, 0, 1, 0, 1), \\
		\mathrm{Ps}_5    &= \mathrm{Ps}(1, 1, 1, 1, 1, 0, 0, 1, 0), &
		\mathrm{Ps}_6    &= \mathrm{Ps}(1, 1, 1, 1, 1, -2, 0, 1, 0), \\
		\mathrm{Ps}_7    &= \mathrm{Ps}(1, 1, 1, 1, 2, 0, 0, 1, 0), &
		\mathrm{Ps}_8    &= \mathrm{Ps}(1, 1, 1, 1, 0, 0, 1, 1, 1), \\
		\mathrm{Ps}_9    &= \mathrm{Ps}(1, 1, 1, 1, 0, 1, 1, 0, 1), &
		\mathrm{Ps}_{10} &= \mathrm{Ps}(1, 1, 1, 1, 1, 0, 0, 1, 1). \\
	\end{aligned}
\end{equation}

Master integrals $\mathrm{Ps}_5$ and $\mathrm{Ps}_6$ form a matrix block which depends on integrals $\mathrm{Ps}_{1,2}$ and enters the right-hand side of equations on master integrals $\mathrm{Ps}_{7,10}$.

In order to build a complete solution for $\mathrm{Ps}_{5,6}$, we need to find a homogeneous solution $\mathbb{H}_{5,6}$.
It appears, however, that in this particular case there is no need to find it, since the corresponding periodic functions $\omega_{5,6}$ are identically 0.
In order to prove this statement, we can consider the asymptotical behavior of the master integrals at $\nu\to\infty$.
The simplest $\mathrm{Ps}_1$ is easily found from \cref{eq:ps-topology} by direct integration:
\begin{equation}
	\label{eq:ps1-sol}
	\mathrm{Ps}_1(\nu) = 2^{6\nu-7} \pi \frac{\Gamma^4(\nu-1)}{\Gamma(3\nu-3) \Gamma(4\nu-4)}.
\end{equation}
Since integrals with all loops cut cannot have ultraviolet divergences,
we can estimate the behavior of the rest of the master integrals at $\nu\to\infty$ from the Baikov representation:
\begin{equation}
	\label{eq:ps-baikov}
	\mathrm{Ps}(1, 1, 1, 1, n_5, \ldots, n_9) \propto
		\int \left( \prod_{i=5}^{9} \frac{\mathrm{d} D_i}{D_i^{n_i}} \right) P(0, 0, 0, 0, D_5, \ldots, D_9)^{\nu - (E+L+1)/2},
\end{equation}
where $P$ is a Baikov polynomial, $E$ is a number of external momenta, and $L$ is a number of loops.
The asymptotics of \cref{eq:ps-baikov} can be found by the saddle-point method, where $P^\nu$ is treated as a rapidly varying function.
The saddle point appears to be determined by
\begin{equation}
	D_5 = D_8 = -\frac43, \quad D_6 = D_7 = D_9 = -1,
\end{equation}
which corresponds to the kinematics where photons momenta point to the vertices of the regular tetrahedron.
Since the rapidly varying function $P^\nu$ stays the same for all master integrals $\mathrm{Ps}_i$,
the asymptotics of any two master integrals can differ at most by some polynomial of $\nu$.
In turn, the inhomogeneous solutions are expressed in terms of sums \eqref{eq:fold-sum}, where $\mathbb{T}_{ij}$ are rational functions of $\nu$.
Thus, their asymptotics at the positive infinity is the same as those of the master integral $\mathrm{Ps}_1$ differing at most by some polynomial of $\nu$.
Both these statements imply the following constraint:
\begin{equation}
	\label{eq:omega-condition}
	\left|\frac{H_k(\nu) \, \Omega_k(\nu)}{\lambda^\nu \,\mathrm{Ps}_1(\nu)}\right| \overset{\nu\to\infty}{\longrightarrow} 0 \ (k=2,\ldots,10)
\end{equation}
for any $\lambda>1$.
The asymptotical behavior of the homogeneous solutions for the sectors with only one master integral ($k=2,3,4,7,\ldots,10$) can be easily found from the scalar equation:
\begin{equation}
	H_k(\nu+1) = M_{kk}(\nu) \, H_k(\nu).
\end{equation}
The behavior of the matrix homogeneous solution $\mathbb{H}_{5,6}$ at $\nu\to\infty$ can be found using the algorithm proposed in Ref.~\cite{Tulyakov:2011},
which is implemented in \DREAM\ in the function \texttt{FindAsymptotics}.
It appears that all homogeneous solutions $H_{2,\ldots,10}(\nu)$ grow exponentially faster than $\mathrm{Ps}_1(\nu)$.
Thus, \cref{eq:omega-condition} holds only if $\Omega_k(\nu) \equiv 0$ for $k=2,\ldots,10$.

After that the master integrals $\mathrm{Ps}_i$ can be easily computed using \DREAM.
Assuming that the package has been already installed, e.g., by the command (see the next section for more details):
\begin{mmaCell}{Input}
	Import["https://bitbucket.org/kmingulov/dream/raw/master/Install.m"];
\end{mmaCell}
we load it with
\begin{mmaCell}[moredefined={DREAM}]{Input}
	<<\,DREAM`
\end{mmaCell}
Then we load the matrix with the dimensional recurrence relations which is contained in the ancillary file \texttt{PsDRR}:
\begin{mmaCell}{Input}
	mat = <<\,"PsDRR"
\end{mmaCell}
\begin{mmaCell}{Output}
	\{\{\mmaFrac{8\,\mmaSup{(-1\,+\,\(\nu\))}{2}}{3\,(-1\,+\,2\(\nu\))\,(-2\,+\,3\(\nu\))\,(-1\,+\,3\(\nu\))\,(-3\,+\,4\(\nu\))\,(-1\,+\,4\(\nu\))},\,0,\,0,\,0,\,0,\,0,\,0,\,0,\,0,\,0\},
	 \{-\mmaFrac{422\,-\,2081\(\nu\)\,+\,3800\mmaSup{\(\nu\)}{2}\,-\,3040\mmaSup{\(\nu\)}{3}\,+\,896 \mmaSup{\(\nu\)}{4}}{24\,(-1\,+\,\(\nu\))\,\mmaSup{(-1\,+\,2\(\nu\))}{4}\,(-2\,+\,3\(\nu\))\,(-3\,+\,4\(\nu\))},\,-\mmaFrac{(-5\,+\,4\(\nu\))\,(-3\,+\,4\(\nu\))}{4(-1\,+\,\(\nu\))\,\mmaSup{(-1\,+\,2\(\nu\))}{4}},\,0,\,0,\,0,\,0,\,0,\,0,\,0,\,0\},\,\ldots\}
\end{mmaCell}
Now we can create all objects necessary for computation of the master integrals with the following command:
\begin{mmaCell}[moredefined={CreateMasters,mat}]{Input}
	CreateMasters["Ps", mat, \mmaUnd{\(\nu\)}];
\end{mmaCell}
Here \texttt{"Ps"} is a prefix which will be used for names of all created objects.
The command will produce the following output:
\begin{mmaCell}[moredefined={Ps1$h}]{Print}
	Diagonal blocks: \{1\}, \{2\}, \{3\}, \{4\}, \{5, 6\}, \{7\}, \{8\}, \{9\}, \{10\}.
	Reducing diagonal blocks…
	Done!
	Block \{1\}
	Block \{2\}
		Ps2 -> Ps1$h: summation direction +
	Block \{3\}
		Ps3 -> Ps1$h: summation direction +
		Ps3 -> Ps2$Ps1$h: summation direction +
		Ps3 -> Ps2$h: summation direction +
	Block \{4\}
		Ps4 -> Ps1$h: summation direction +
	Block \{5, 6\}
		Ps5 -> Ps1$h: summation direction +
		Ps5 -> Ps2$Ps1$h: summation direction +
	...
\end{mmaCell}
The output shows which matrix blocks were found and summation direction used in each encountered sum of form \eqref{eq:fold-sum-representation}.
\texttt{Ps1\$h}, \texttt{Ps2\$Ps1\$h} (and so on) are the names of the created objects.
Objects are represented as \Mathematica\ functions, their first argument being the name of the method and the rest being the method arguments.
The general interface to the user is provided via the object \texttt{Ps}.
When \DREAM\ constructs the object hierarchy above, it sets all homogeneous solutions to zero by default.
Therefore we set the homogeneous solution for $\mathrm{Ps}_1$, \cref{eq:ps1-sol}, by the command:
\begin{mmaCell}[moredefined={Ps,Ps1,mSetSolution}]{Input}
	Ps[mSetSolution, \{Ps1 \(\to\) \mmaFrac{\mmaSup{2}{6\mmaUnd{\(\nu\)}\,-\,7}\,\(\pi\)\,\mmaSup{Gamma[\mmaUnd{\(\nu\)}\,-\,1]}{4}}{Gamma[3\mmaUnd{\(\nu\)}\,-\,3]\,Gamma[4\mmaUnd{\(\nu\)}\,-\,4]}\}, \mmaUnd{\(\nu\)}];
\end{mmaCell}
With these two commands, we are already prepared to calculate the integrals.
We simply execute the following command:
\begin{mmaCell}[moredefined={Ps,mEvaluate}]{Input}
	Ps[mEvaluate, All, 2\,-\,\mmaUnd{\(\epsilon\)}, 1, 20]
\end{mmaCell}
The method \texttt{mEvaluate}, that we used, has the following arguments:
keyword \texttt{All} indicating that all master integrals must be computed,
the value of the variable $\nu=2-\epsilon$,
the order of the expansion by $\epsilon$,
and required precision of the result.
The output of this command looks like:
\begin{mmaCell}{Output}
	\{Ps1[2\,-\,\(\epsilon\)] \(\to\) 8.37758040957278197\,+\,49.7863121292874579\,\(\epsilon\)\,+\,\mmaSup{O[\(\epsilon\)]}{2},
	 Ps2[2\,-\,\(\epsilon\)] \(\to\) -6.69283516277769869\,-\,40.3817198468457380\,\(\epsilon\)\,+\,\mmaSup{O[\(\epsilon\)]}{2},
	 Ps3[2\,-\,\(\epsilon\)] \(\to\) 10.33542556009994006\,+\,70.4450919445169533\,\(\epsilon\)\,+\,\mmaSup{O[\(\epsilon\)]}{2},
	 Ps4[2\,-\,\(\epsilon\)] \(\to\) 16.20896101168141433\,+\,113.953031911002724\,\(\epsilon\)\,+\,\mmaSup{O[\(\epsilon\)]}{2},
	 Ps5[2\,-\,\(\epsilon\)] \(\to\) 5.16274120289082745\,+\,31.3277288470935572\,\(\epsilon\)\,+\,\mmaSup{O[\(\epsilon\)]}{2},
	 Ps5s1[2\,-\,\(\epsilon\)] \(\to\) 0.00621507125236686853\,+\,0.0457018952249140321\,\(\epsilon\)\,+\,\mmaSup{O[\(\epsilon\)]}{2},
	 Ps6[2\,-\,\(\epsilon\)] \(\to\) 6.255140817819126671\,+\,38.9343044807864777\,\(\epsilon\)\,+\,\mmaSup{O[\(\epsilon\)]}{2},
	 Ps7[2\,-\,\(\epsilon\)] \(\to\) -4.33242115540336673\,-\,27.1899959149171734\,\(\epsilon\)\,+\,\mmaSup{O[\(\epsilon\)]}{2},
	 Ps8[2\,-\,\(\epsilon\)] \(\to\) -12.75082019938672722\,-\,90.587556664362580\,\(\epsilon\)\,+\,\mmaSup{O[\(\epsilon\)]}{2},
	 Ps9[2\,-\,\(\epsilon\)] \(\to\) -17.9929365915101837\,-\,129.054599478278968\,\(\epsilon\)\,+\,\mmaSup{O[\(\epsilon\)]}{2},
	 Ps10[2\,-\,\(\epsilon\)] \(\to\) -8.05018524974869544\,-\,55.425826999342817\,\(\epsilon\)\,+\,\mmaSup{O[\(\epsilon\)]}{2}\}
\end{mmaCell}
The master integrals were computed with precision of 1000 digits in less than a minute on a laptop with the four-core processor Intel i7-3630QM.

Using these values, we can obtain the expression for $\Gamma^{\text{LO}}(\mathrm{Ps}\to4\gamma)$:
\begin{multline}
	\Gamma^{\text{LO}}(\mathrm{Ps}\to4\gamma) = 
		\frac{m \alpha^5}{2} \left(\frac{\alpha}{\pi}\right)^2 \times 0.274292031337256574490596323020\ldots \\ \overset{(1000)}{=}
		\frac{m \alpha^5}{2} \left(\frac{\alpha}{\pi}\right)^2 \left( \frac{112}{5}-\frac{3 \pi ^2}{2}+\frac{3\,\zeta_3}{10}+\frac{\pi ^4}{24}+\frac{697}{15} \pi ^2 \log 2-\frac{152}{5} \pi ^2 \log 3 \right).
\end{multline}
The expression in terms of transcendental constants was recovered by means of integer relation algorithm \texttt{PSLQ} \cite{FergusonBailey:1992:PSLQ}
and obtained here for the first time to the best of our knowledge.
Our result is in the perfect numerical agreement with that of Adkins and Pfahl.

\subsection{Four-loop massive tadpole diagrams}

\mmaResetCellIndex

\begin{figure}[t]
	\centering
	\setlength{\tabcolsep}{5pt}
	\newcommand{\miscale}{0.4}
	\newcommand{\misize}{\footnotesize}
	\begin{tabular}{ccccc}
		\includegraphics[scale=\miscale,trim={26pt 37pt 18pt 36pt},clip=true]{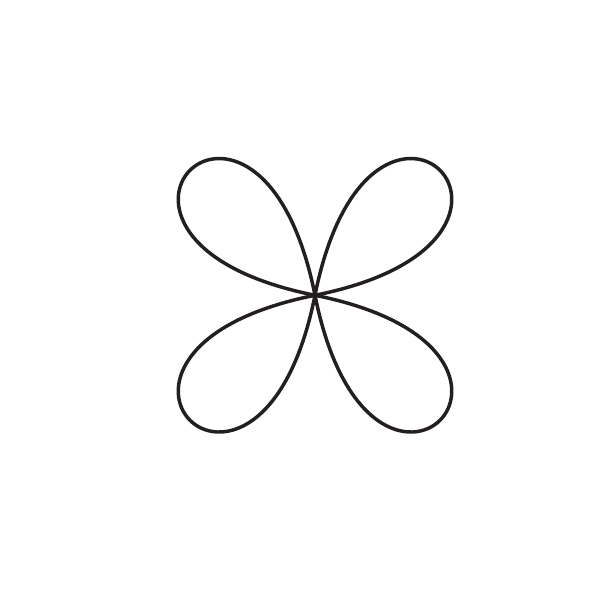} &
		\includegraphics[scale=\miscale,trim={26pt 37pt 18pt 36pt},clip=true]{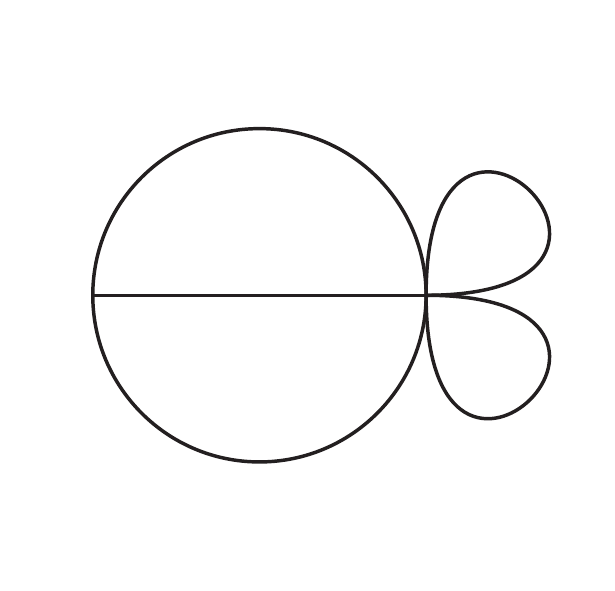} &
		\includegraphics[scale=\miscale,trim={26pt 37pt 18pt 36pt},clip=true]{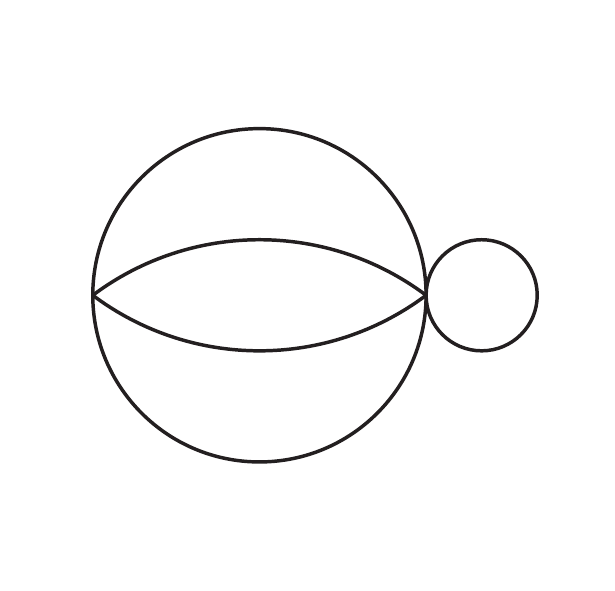} &
		\includegraphics[scale=\miscale,trim={26pt 37pt 18pt 36pt},clip=true]{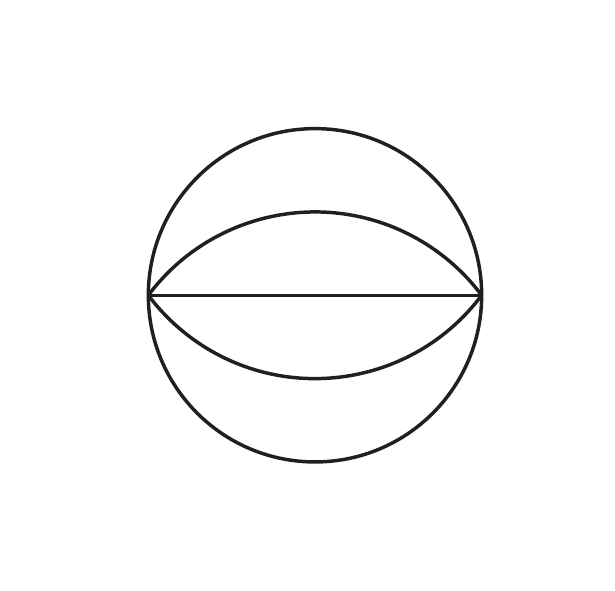} &
		\includegraphics[scale=\miscale,trim={26pt 37pt 18pt 36pt},clip=true]{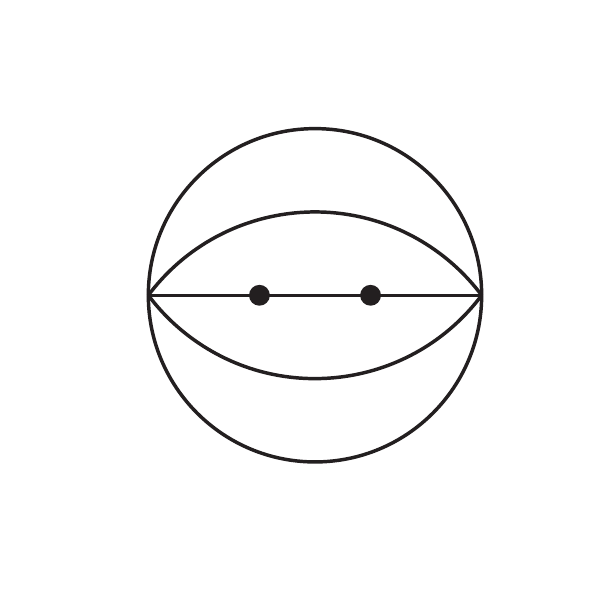} \\
		\misize $T_1$ & \misize $T_2$ & \misize $T_3$ & \misize $T_4$ & \misize $T_5$ 
		\vspace{1em}\\
		\includegraphics[scale=\miscale,trim={26pt 37pt 18pt 36pt},clip=true]{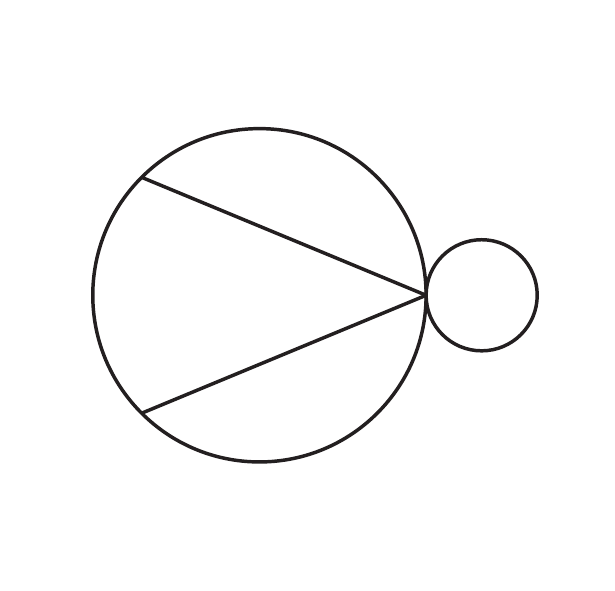} &
		\includegraphics[scale=\miscale,trim={26pt 37pt 18pt 36pt},clip=true]{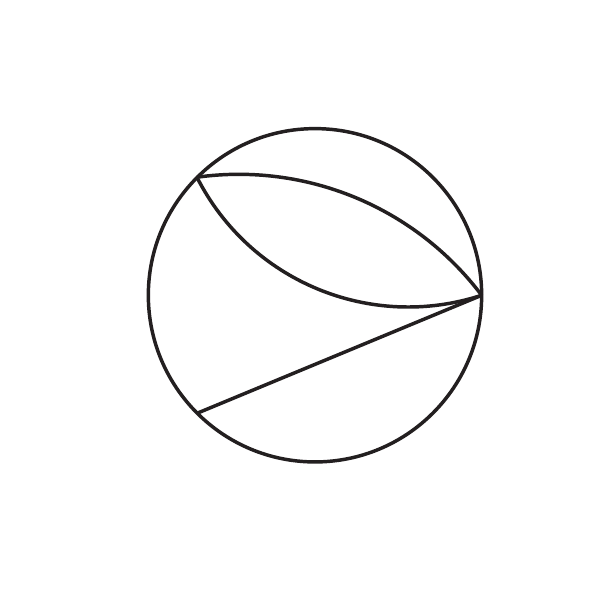} &
		\includegraphics[scale=\miscale,trim={26pt 37pt 18pt 36pt},clip=true]{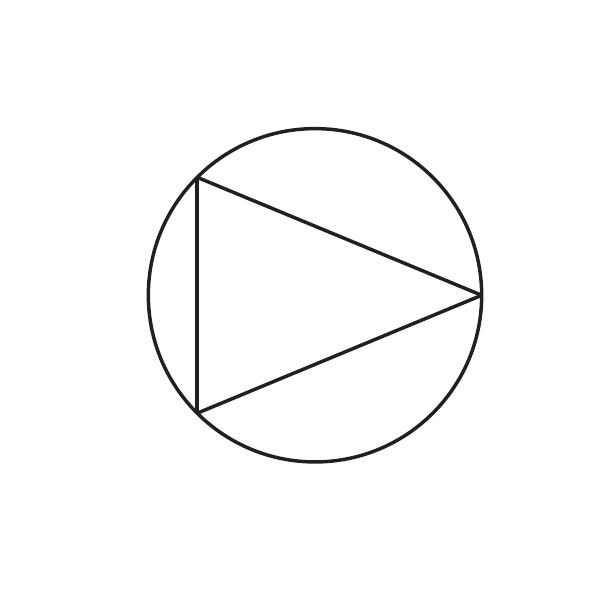} &
		\includegraphics[scale=\miscale,trim={26pt 37pt 18pt 36pt},clip=true]{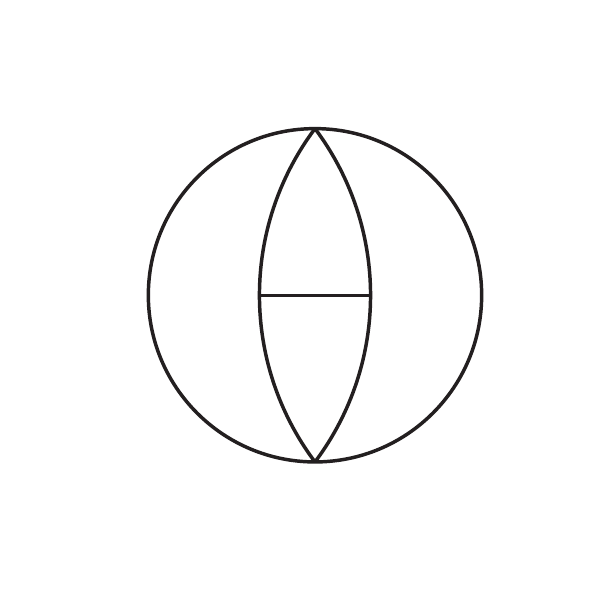} &
		\includegraphics[scale=\miscale,trim={26pt 37pt 18pt 36pt},clip=true]{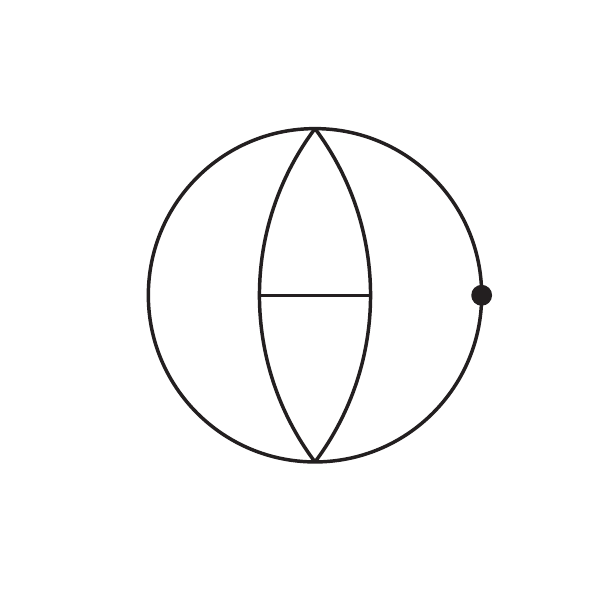} \\
		\misize $T_6$ & \misize $T_7$ & \misize $T_8$ & \misize $T_9$ & \misize $T_{10}$
	\end{tabular} 
	\setlength{\tabcolsep}{6pt}
	\renewcommand{\arraystretch}{1.0}
	\caption{Master-integrals for the four-loop massive tadpole topology.}
	\label{pic:tadpole-masters}	
	\vspace{5mm} 	    
\end{figure}

Let us now move to a more complex example, four-loop fully massive tadpoles of cat-eye topology:
\begin{align}
	\label{eq:tadpole-topo}
	T(n_1, n_2, n_3, n_4, n_5, n_6, n_7) &= 
		\int \frac{1}{\pi^{4d/2}} \frac{\prod_{i=1}^{4} \mathrm{d}l_i}{\prod_{j=1}^{7} D_j^{n_j}}.
	\\
	T(1, 1, 1, 1, 1, 1, 1) &= \raisebox{-0.45\height}{\includegraphics[scale=0.4,trim={42pt 37pt 38pt 36pt},clip=true]{T9}}.
\end{align}
Denominators $D_j$ can be chosen as:
\begin{equation}
\begin{aligned}
	D_{1,2,3,4} &= l_{1,2,3,4}^2-1, & 
	D_{5} &= (l_3-l_4)^2-1, \\
	D_{6} &= (l_1+l_2+l_3)^2-1, &
	D_{7} &= (l_2+l_3-l_4)^2-1.
\end{aligned}
\end{equation}%
Performing the IBP-reduction with \LiteRed, we arrive at ten master integrals shown on Fig.~\ref{pic:tadpole-masters}.
The difference equation system has two nontrivial matrix blocks: $T_{4,5}$ and $T_{9,10}$;
what makes this example even more difficult, is that $T_{9,10}$-block depends on $T_{4,5}$-block.

For further convenience, we pass from master integrals $T_i$ to:
\begin{equation}
	\widetilde{T}_i(\nu) = \frac{ T_i(\nu) }{ T_1(\nu) }
\end{equation}
and omit the tilde in what follows.

In the previous example, we could proceed without knowing the homogeneous solution corresponding to the $\mathrm{Ps}_{5,6}$-block.
This is not the case for the present master integral topology and we are obliged to construct these solutions for two nontrivial matrix blocks.
Both blocks can be reduced to a $2$-nd order recurrence relation by $\nu$
and their solution is described in the parallel paper \cite{LeeMingulov:Meromorphic}.
Here we will restrict ourselves to the question of how to embed these solutions into the \DREAM\ objects hierarchy.

The two independent homogeneous solutions for $T_{4,5}$-block given in \cite{LeeMingulov:Meromorphic} read as follows:
\begin{equation}
	\label{eq:tadpole-homo}
	\begin{aligned}
		H_{41}(\nu) &= f_{41}(\nu) \, S_4(\nu), &
		H_{42}(\nu) &= f_{42}(\nu) \, S_4(2-\nu),
	\end{aligned}
\end{equation}
where
\begin{equation}
	\label{eq:tadpole-f}
	\begin{aligned}
		S_4(\nu) &= \sum_{n=0}^\infty \left(\frac{\left(\frac{25}{8}\right)^{-n-\frac{5}{4}} a_n}{\nu -n-\frac{5}{4}}+C_4\,\frac{\left(\frac{25}{8}\right)^{-n-\frac{7}{4}} b_n}{\nu -n-\frac{7}{4}}\right),
		\\
		f_{41}(\nu) &= \left(\frac{125}{64}\right)^\nu \frac{ \Gamma(2-2\nu) \, \Gamma\left(\frac{4}{3}-\nu\right) \Gamma\left(\frac{5}{3}-\nu\right) }{ \Gamma^4(1-\nu) },
		\\
		f_{42}(\nu) &= \frac{625}{64} \frac{ \Gamma(2-2\nu) \, \Gamma\left(\frac{7}{4}-\nu\right)}{\Gamma^4(1-\nu) \, \Gamma\left(\nu-\frac{1}{4}\right)},
		\\
		C_4 &= -7 \sqrt{\frac{5}{6}} \frac{\Gamma \left(\frac{3}{4}\right)^2}{\Gamma \left(\frac{1}{4}\right)^2} = -0.729983187537670\ldots
	\end{aligned}
\end{equation}
and $a_n$ and $b_n$ are defined recursively:
\begin{equation}
	\label{eq:tadpole-recur}
	\begin{aligned}
		a_{-1} &= 0, & a_0 &= 1, & 
		a_{n\geq 1} &= \frac{\left(8784 n^2-13176 n+5197\right)}{2880 n \, (2 n-1)} \, a_{n-1} +
			\frac{5 (12 n-17) (12 n-13)}{72 n \, (2 n-1)} \, a_{n-2},
		\\
		b_{-1} &= 0, & b_0 &= 1, & 
		b_{n\geq 1}&=\frac{\left(8784 n^2-4392 n+805\right)}{2880 n \, (2 n+1)} \, b_{n-1} +
			\frac{5 (12 n-11) (12 n-7)}{72 n \, (2 n+1)} \, b_{n-2}.
	\end{aligned}
\end{equation}
The homogeneous solution $H_{41}(\nu)$ can be implemented as a \Mathematica\ function in the following way:
\newcommand{\pbl}{\(\pmb{\big[}\)}
\newcommand{\pbr}{\(\pmb{\big]}\)}
\newcommand{\ppl}{\(\pmb{\big(}\)}
\newcommand{\ppr}{\(\pmb{\big)}\)}
\begin{mmaCell}[moredefined={T4homo1, ConvergenceAccelerate},morepattern={o_Integer, p_Integer, p, o, \#1, \#},morelocal={n, res, a, b, exp1, exp2, sd1, sd2, Nt, C4, oc, f41, n_Integer, n_}]{Input}
	T4homo1\pbl\mmaPat{\(\nu\)_}, \{\mmaPat{\(\epsilon\)_}, o_Integer\}, p_Integer\pbr\,:= 
	  Module\pbl\{
	      (* LOCAL VARIABLES *) 
	      n, \mmaLoc{\(\nu\)0}\,=\,\mmaPat{\(\nu\)}\,/.\,\mmaPat{\(\epsilon\)}\,\(\to\)\,0, oc\,=\,o, f41, C4\,=\,-7\mmaSqrt{\mmaFrac{5}{6}}\mmaFrac{\mmaSup{Gamma[\mmaFrac{3}{4}]}{2}}{\mmaSup{Gamma[\mmaFrac{1}{4}]}{2}}, res\,=\,\{0\}, a, b,
	      (* Functions for denominators \mmaSup{(\(\nu\)-n-5/4)}{-1} and \mmaSup{(\(\nu\)-n-7/4)}{-1}. *)
	      exp1, exp2, sd1, sd2,
	      (* Number of computed terms, which will be used to estimate 
	         the value of the sum. *)
	      Nt\,=\,6.9p\,+\,20
	    \},
	
	    (* Function \mmaSub{f}{41} has singularities or zeros at some values of \(\nu\).
	       We adjust expansion order, so that we will obtain the expression 
	       with terms by \(\epsilon\) up to order o after multiplication by \mmaSub{f}{41}. *)
	    If\pbl\mmaLoc{\(\nu\)0}\,\(\in\)\,Reals && \mmaLoc{\(\nu\)0}\,\(\geq\)\,1 && MatchQ\pbl\vphantom{0}FractionalPart[\mmaLoc{\(\nu\)0}],\,\mmaFrac{1}{2}|\mmaFrac{1}{3}|\mmaFrac{2}{3}\pbr, oc++\pbr;
	    If\pbl\mmaLoc{\(\nu\)0}\,\(\in\)\,Reals && \mmaLoc{\(\nu\)0}\,\(\geq\)\,1 && MatchQ\pbl\vphantom{0}FractionalPart[\mmaLoc{\(\nu\)0}],\,0\pbr, oc\,-=\,3\pbr;
	
	    (* \mmaSub{a}{n} and \mmaSub{b}{n}. *)
	    a[-1]\,=\,b[-1]\,=\,0;
	    a[0]\,=\,N\pbl\mmaSup{\bigg(\mmaFrac{25}{8}\bigg)}{-\mmaFrac{5}{4}}, Floor[2p]\pbr; b[0]\,=\,N\pbl\mmaSup{\bigg(\mmaFrac{25}{8}\bigg)}{-\mmaFrac{7}{4}}C4, Floor[2p]\pbr;
	    a[n_Integer]:=\,a[n]\,=\,\mmaFrac{(8784\mmaSup{n}{2}\,-\,13176n\,+\,5197)\,a[n\,-\,1]}{2880\,n\,(2n\,-\,1)}\,+\,\mmaFrac{5\,(12n\,-\,17)\,(12n\,-\,13)\,a[n\,-\,2]}{72\,n\,(2n\,-\,1)};
	    b[n_Integer]:=\,b[n]\,=\,\mmaFrac{(8784\mmaSup{n}{2}\,-\,4392n\,+\,805)\,b[n\,-\,1]}{2880\,n\,(2n\,+\,1)}\,+\,\mmaFrac{5\,(12n\,-\,11)\,(12n\,-\,7)\,b[n\,-\,2]}{72\,n\,(2n\,+\,1)};
	
	    (* Denominators. *)
	    sd1\,=\,Series\pbl\mmaFrac{1}{\mmaPat{\(\nu\)}\,-\,n\,-\,\mmaFrac{5}{4}}, \{\mmaPat{\(\epsilon\)},\,0,\,oc\}\pbr; sd2\,=\,Series\pbl\mmaFrac{1}{\mmaPat{\(\nu\)}\,-\,n\,-\,\mmaFrac{7}{4}}, \{\mmaPat{\(\epsilon\)},\,0,\,oc\}\pbr;
	    exp1[n_]:=\,If\pbl\mmaLoc{\(\nu\)0}\,-\,n\,-\,\mmaFrac{5}{4}\,===\,0, Series\pbl\mmaFrac{1}{\mmaPat{\(\nu\)}\,-\,n\,-\,\mmaFrac{5}{4}}, \{\mmaPat{\(\epsilon\)},\,0,\,oc\}\pbr, sd1\pbr;
	    exp2[n_]:=\,If\pbl\mmaLoc{\(\nu\)0}\,-\,n\,-\,\mmaFrac{7}{4}\,===\,0, Series\pbl\mmaFrac{1}{\mmaPat{\(\nu\)}\,-\,n\,-\,\mmaFrac{7}{4}}, \{\mmaPat{\(\epsilon\)},\,0,\,oc\}\pbr, sd2\pbr;
	
	    (* Computing accumulative sums. *)
	    Do\pbl\vphantom{0}AppendTo\pbl\vphantom{0}res, Last[res]+\mmaSup{\bigg(\mmaFrac{25}{8}\bigg)}{-\mmaFnc{n}}(a[\mmaFnc{n}]exp1[\mmaFnc{n}]\,+\,b[\mmaFnc{n}]exp2[\mmaFnc{n}])\pbr,
	      \{\mmaFnc{n},\,0,\,Nt\}\pbr;
	
	    (* Performing convergence acceleration by means of DREAM *)
	    (* function ConvergenceAccelerate and returning the result. *)
	    f41\,=\,\mmaSup{\bigg(\mmaFrac{125}{64}\bigg)}{\mmaLoc{\(\nu\)}}\mmaFrac{Gamma[2\,-\,2\mmaLoc{\(\nu\)}]\,Gamma[\mmaFrac{4}{3}\,-\,\mmaLoc{\(\nu\)}]\,Gamma[\mmaFrac{5}{3}\,-\,\mmaLoc{\(\nu\)}]}{\mmaSup{Gamma[1-\mmaLoc{\(\nu\)}]}{4}};
	    f41\,N\pbl\vphantom{0}ConvergenceAccelerate\pbl\vphantom{0}res, Floor\pbl0.07Nt\pbr, \{1,\,2\}\pbr, p\pbr
	  \pbr;
\end{mmaCell}
The second homogeneous solution $H_{42}(\nu)$ and the solutions for $T_{9,10}$-block are implemented in much the same way and are presented in the ancillary file \texttt{Examples.nb}.

Taking into account that the inhomogeneous solutions can be created by \DREAM,
we can proceed with the DRA method and fix the periodic functions $\bm{\Omega}_4(\nu)$ and $\bm{\Omega}_9(\nu)$ as described in the parallel paper \cite{LeeMingulov:Meromorphic}.
The periodic functions read as follows:
\begin{alignat}{2}
	\omega_{41}(\nu)&=\frac{16}{5 \sqrt{\pi }}\left(\frac{2}{15}\right)^{1/4} \Gamma \left(\tfrac{1}{4}\right)^2, \qquad & \omega_{42}(\nu)&=0, \\
	\omega_{91}(\nu)&=-\frac{8 \pi ^2}{3 \, \Gamma \left(\frac{1}{3}\right)}, & \omega_{92}(\nu)&=0.
\end{alignat}

Now we can embed the homogeneous solutions into \DREAM\ objects hierarchy.
First, we create the objects which can be used by \DREAM\ for the computation of these homogeneous solutions.
\begin{mmaCell}[moredefined={ONew,TFunction,T4homo1,T$h}, morepattern={o, p}]{Input}
	ONew[T4$hf, TFunction, Function\pbl\{\mmaPat{\(\nu\)}, \mmaPat{\(\epsilon\)}, o, p\},
	  \mmaFrac{16}{5\mmaSqrt{\mmaDef{\(\pi\)}}}\mmaSup{\bigg(\mmaFrac{2}{15}\bigg)}{\!1/4}\mmaSup{Gamma\big[\mmaFrac{1}{4}\big]}{2}T4homo1[\mmaPat{\(\nu\)}, \{\mmaPat{\(\epsilon\)},\,o\}, p]\pbr, \big\{\mmaFrac{125}{256},\,-\mmaFrac{1}{2},\,0\big\}];
\end{mmaCell}
\begin{mmaCell}[moredefined={ONew,TFunction,T9homo1,T9homo2,T$h}, morepattern={o, p}]{Input}
	ONew[T9$hf, TFunction, Function\pbl\{\mmaPat{\(\nu\)}, \mmaPat{\(\epsilon\)}, o, p\},
	  -\mmaFrac{8\,\mmaSup{\mmaDef{\(\pi\)}}{2}}{3\,Gamma[\mmaFrac{1}{3}]}\,T9homo1[\mmaPat{\(\nu\)}, \{\mmaPat{\(\epsilon\)},\,o\}, p]\pbr, \big\{\mmaFrac{35\,+\,13\mmaSqrt{13}}{216},\,-\mmaFrac{3}{2},\,0\big\}];
\end{mmaCell}
The function \texttt{ONew} is the tool for creating new objects in \DREAM.
Its first argument is the name of the object,
the second one is the object's class (namely \TFunction, a special class which allows to embed custom functions into the objects hierarchy),
and the left arguments are the constructor arguments:
a function of four arguments and its asymptotics on the positive and negative infinity by $\nu$ in terms of three parameters
\begin{equation}
	\{q, \alpha, \beta\} \Leftrightarrow f(\nu) \overset{\nu\to\pm\infty}{\sim} \frac{q^\nu}{\nu^\alpha \, \Gamma(\nu)^\beta}.
\end{equation}
The list $\{q, \alpha, \beta\}$ can be found either from the explicit expressions of the homogeneous solutions in \cref{eq:tadpole-homo,eq:tadpole-f,eq:tadpole-recur} or from the second-order recurrence relation with help of \DREAM\ function \texttt{FindAsymptotics}.

We could embed the object \texttt{T4\$hf} into the objects hierarchy right away.
However since there are the master integrals depending on $T_{4,5}$-block which would require to compute the function \texttt{T4homo1} at many consequent points, it appears to be useful to organize this calculation with the help of the recurrence relation.
Therefore, we create an auxiliary object of class \TRecurrence:
\begin{mmaCell}[moredefined={ONew,TRecurrence,T4$hf,T$h}]{Input}
	ONew[T4$h, TRecurrence, T4$hf, \big\{1, -\mmaFrac{\mmaSup{(\mmaUnd{\(\nu\)}\,+\,1)}{3}(549\mmaSup{\mmaUnd{\(\nu\)}}{2}\,+\,16)}{8(2\mmaUnd{\(\nu\)}\,+\,1)(3\mmaUnd{\(\nu\)}\,+\,1)(3\mmaUnd{\(\nu\)}\,+\,2)(4\mmaUnd{\(\nu\)}\,+\,1)(4\mmaUnd{\(\nu\)}\,+\,3)},
	  -\mmaFrac{1125(\mmaSup{\mmaUnd{\(\nu\)}}{6}\,+\,3\mmaSup{\mmaUnd{\(\nu\)}}{5}\,+\,3\mmaSup{\mmaUnd{\(\nu\)}}{4}\,+\,\mmaSup{\mmaUnd{\(\nu\)}}{3})}{16(2\mmaUnd{\(\nu\)}\,-\,1)(2\mmaUnd{\(\nu\)}\,+\,1)(3\mmaUnd{\(\nu\)}\,+\,1)(3\mmaUnd{\(\nu\)}\,+\,2)(4\mmaUnd{\(\nu\)}\,+\,1)(4\mmaUnd{\(\nu\)}\,+\,3)}, 0\big\}, \mmaUnd{\(\nu\)},
	  \big\{\mmaFrac{125}{256},\,-\mmaFrac{1}{2},\,0\big\}, \big\{\mmaFrac{125}{256},\,-\mmaFrac{1}{2},\,0\big\}, \mmaFrac{3}{10}];
\end{mmaCell}
Here two lists $\{\frac{125}{256}, -\frac12, 0\}$ denote the asymptotics of the object on the positive and the negative infinity, 
and fraction $\frac{3}{10}$ means that the working precision will be increased by $\frac{3 p_r}{10}$ for each computed term (where $p_r$ is the required precision).
This is necessary, since using a recurrence relation leads to precision loss and we can avoid it by increasing precision in advance.

The final step is to create the \DREAM\ objects hierarchy.
We load the matrix with recurrence relations from file \texttt{tDRR} (note that we multiply it by $\nu^4$ which corresponds to the division of all master integrals by $T_1(\nu)=\Gamma^4(1-\nu)$)
and create the hierarchy with the following commands:
\begin{mmaCell}[moredefined={CreateMasters,mat,T4$h,T9$hf}]{Input}
	mat = <<\,"tDRR";
	mat = \mmaSup{\mmaUnd{\(\nu\)}}{4}\,mat;
	CreateMasters["T", mat, \mmaUnd{\(\nu\)}, Append \(\to\) \{T4 \(\to\) T4$h, T9 \(\to\) T9$hf\}];
\end{mmaCell}
The used option \texttt{Append} specifies objects which should be ``appended'' to other objects constituting the specified integral.

The periodic function for the scalar master integrals $T_{1,2,3,6,7,8}$ are fixed with the regular DRA method and can be set in a program in the following way:
\begin{mmaCell}[moredefined={T,T1,T2,T3,T6,T7,T8,mSetSolution}]{Input}
	T[mSetSolution, \{T1 \(\to\) 1, T2 \(\to\) -\mmaFrac{8\mmaSqrt{\(\pi\)}}{3\mmaSqrt{3}}\mmaSup{\bigg(\mmaFrac{3}{4}\bigg)}{\mmaUnd{\(\nu\)}}\mmaFrac{Gamma[\mmaFrac{3}{2}\,-\,\mmaUnd{\(\nu\)}]}{Gamma[1\,-\,\mmaUnd{\(\nu\)}]},
	  T3 \(\to\) -\mmaFrac{27\mmaSqrt{3\mmaDef{\(\pi\)}}}{32}\mmaSup{\bigg(\mmaFrac{16}{27}\bigg)}{\mmaUnd{\(\nu\)}}\mmaFrac{Gamma[\mmaFrac{4}{3}\,-\,\mmaUnd{\(\nu\)}]\,Gamma[\mmaFrac{3}{2}\,-\,\mmaUnd{\(\nu\)}]\,Gamma[\mmaFrac{5}{3}\,-\,\mmaUnd{\(\nu\)}]}{Gamma[2\,-\,\mmaUnd{\(\nu\)}]\,\mmaSup{Gamma[1\,-\,\mmaUnd{\(\nu\)}]}{2}},
	  T6 \(\to\) \mmaFrac{64\,\mmaDef{\(\pi\)}}{27}\mmaSup{\bigg(\mmaFrac{9}{16}\bigg)}{\mmaUnd{\(\nu\)}}\mmaFrac{\mmaSup{Gamma[\mmaFrac{3}{2}\,-\,\mmaUnd{\(\nu\)}]}{2}}{\mmaSup{Gamma[1\,-\,\mmaUnd{\(\nu\)}]}{2}}, T7 \(\to\) \mmaFrac{9\,\mmaDef{\(\pi\)}}{4}\mmaSup{\bigg(\mmaFrac{4}{9}\bigg)}{\mmaUnd{\(\nu\)}}\mmaFrac{Gamma[\mmaFrac{4}{3}\,-\,\mmaUnd{\(\nu\)}]\,\mmaSup{Gamma[\mmaFrac{3}{2}\,-\,\mmaUnd{\(\nu\)}]}{2}\,Gamma[\mmaFrac{5}{3}\,-\,\mmaUnd{\(\nu\)}]}{Gamma[2\,-\,\mmaUnd{\(\nu\)}]\,\mmaSup{Gamma[1\,-\,\mmaUnd{\(\nu\)}]}{3}},
	  T8 \(\to\) \mmaFrac{128\mmaSqrt{2}\mmaDef{\(\pi\)}}{81}\mmaSup{\bigg(\mmaFrac{27}{64}\bigg)}{\mmaUnd{\(\nu\)}}\mmaFrac{\mmaSup{Gamma[\mmaFrac{3}{2}\,-\,\mmaUnd{\(\nu\)}]}{3}\,Gamma[\mmaFrac{7}{4}\,-\,\mmaUnd{\(\nu\)}]\,Gamma[2\,-\,\mmaUnd{\(\nu\)}]\,Gamma[\mmaFrac{9}{4}\,-\,\mmaUnd{\(\nu\)}]}{\mmaSup{Gamma[1\,-\,\mmaUnd{\(\nu\)}]}{3}\,Gamma[\mmaFrac{11}{6}\,-\,\mmaUnd{\(\nu\)}]\,Gamma[\mmaFrac{13}{6}\,-\,\mmaUnd{\(\nu\)}]\,Gamma[\mmaFrac{5}{2}\,-\,\mmaUnd{\(\nu\)}]}\}, \mmaUnd{\(\nu\)}];
\end{mmaCell}
Now we can compute the master integrals $T_{1, \ldots, 10}$ with already mentioned method \texttt{mEvaluate}:
\begin{mmaCell}[moredefined={T,mEvaluate}]{Input}
	T[mEvaluate, All, 2\,-\,\mmaUnd{\(\epsilon\)}, 7, 500]
\end{mmaCell}
The running time of the command above was about 20 minutes on a laptop with four-core processor Intel i7-6700HQ.

Values of some four-loop fully massive tadpoles obtained with \DREAM\ are presented in the appendix.

\section{Brief review of \DREAM\ functionality}
\label{sec:dream-manual}

\mmaResetCellIndex

In this section we briefly review some functionality of the \DREAM\ package.
As already mentioned above, the latest version of the package can be installed by:
\begin{mmaCell}{Input}
	Import["https://bitbucket.org/kmingulov/dream/raw/master/Install.m"];
\end{mmaCell}
Afterwards the package can be loaded by the following command:
\begin{mmaCell}[moredefined={DREAM}]{Input}
	<<\,DREAM`
\end{mmaCell}
\DREAM\ has an extensive documentation, which is accessible via standard \Mathematica\ Documentation Center (to open it, type in, e.g., \texttt{DREAM/guide/DREAMPackage} in the address bar).
The documentation describes behavior of all functions and classes in detail.
In this section, we will only cover some of them. 

\subsection{\texttt{CreateMaster} and \texttt{CreateMasters}}

\texttt{CreateMaster} and \texttt{CreateMasters} are procedures which automatically create all objects necessary for computation of master integrals.
The function \texttt{CreateMasters}, which was used in the examples above, has the following form:
\begin{mmaCell}[moredefined={CreateMasters}]{Input}
	CreateMasters["name", mat, \mmaUnd{\(\nu\)}]
\end{mmaCell}
Here \texttt{"name"} is a prefix which will be used for all created objects and \texttt{mat} is a matrix $\mathbb{M}(\nu)$ of lowering dimensional recurrence relations.
If there are nontrivial matrix blocks, this function passes from block master integrals $J_1(\nu), \ldots, J_n(\nu)$ to ``shifted'' master integrals $J_1(\nu), J_1(\nu+1), \ldots, J_1(\nu+n-1)$.

Additionally to \texttt{CreateMasters}, there is a function \texttt{CreateMaster} which constructs objects for one specific master integral by a first-order recurrence relation:
\begin{mmaCell}[moredefined={CreateMaster}]{Input}
	CreateMaster[J, \{\mmaSub{c}{0}, \{\mmaSub{c}{1},\,\mmaSub{m}{1}\}, \{\mmaSub{c}{2},\,\mmaSub{m}{2}\}, \ldots\}, \mmaUnd{\(\nu\)}]
\end{mmaCell}
by $n$-th order recurrence relation:
\begin{mmaCell}[moredefined={CreateMaster,n}]{Input}
	CreateMaster[J, \{\mmaSub{c}{01}, \mmaSub{c}{02}, \ldots, \mmaSub{c}{0n}, \{\mmaSub{c}{1},\,\mmaSub{m}{1}\}, \{\mmaSub{c}{2},\,\mmaSub{m}{2}\}, \ldots\}, \mmaUnd{\(\nu\)}]
\end{mmaCell}
or by its expression in terms of other master integrals:
\begin{mmaCell}[moredefined={CreateMaster}]{Input}
	CreateMaster[J, \{\{\mmaSub{c}{1},\,\mmaSub{m}{1}\}, \{\mmaSub{c}{2},\,\mmaSub{m}{2}\}, \ldots\}, \mmaUnd{\(\nu\)}]
\end{mmaCell}
Here \texttt{J} is a symbol, which will be used for the new master integral, $c_i$ are rational functions of $\nu$, and $m_i$ are already created master integrals.

Both \texttt{CreateMasters} and \texttt{CreateMaster} has an option \texttt{Append} allowing one to add  custom objects to a master integral:
\begin{mmaCell}[moredefined={CreateMaster}]{Input}
	CreateMaster[J, \{\ldots\}, \mmaUnd{\(\nu\)}, Append \(\to\) \{obj1, obj2, \ldots\}]
\end{mmaCell}
\begin{mmaCell}[moredefined={CreateMasters}]{Input}
	CreateMasters["J", mat, \mmaUnd{\(\nu\)}, Append \(\to\) \{J1 \(\to\) \{obj11, obj12, \ldots\}, 
	                                      J2 \(\to\) \{obj21, obj22, \ldots\}, \ldots\}]
\end{mmaCell}

\subsection{\TMasters\ class}

\TMasters\ is a class representing a topology of master integrals; 
objects of this class are created by the function \texttt{CreateMasters} (in the examples above, objects \texttt{Ps} and \texttt{T} were instances of \TMasters).
The most useful methods of \TMasters\ are:
\begin{description}
	\item[\texttt{mEvaluate(mis, $\nu$, o, p)}]
		Computes the master integrals specified in the list \texttt{mis} at $\nu$ (for example, $2-\epsilon$) with expansion order \texttt{o} and precision \texttt{p}.
		\texttt{mis} can be \texttt{All}, meaning that all master integrals in the topology must be computed.

	\item[\texttt{mPurge()}]
		Completely removes all objects used by this integral topology.

	\item[\texttt{mRangeEvaluate(mis, $\nu$, o, n, p)}]
		Computes the master integrals from the list \texttt{mis} at \texttt{n} consequent points.
		This method returns values at $\nu, \nu\pm1, \ldots, \nu+n$ (depending on the sign of $n$).

	\item[\texttt{mSave(file)}]
		Saves all objects to \texttt{file}.
		The file can be later loaded by the function \texttt{OLoad}.

	\item[\texttt{mSetSolution(sols, $\nu$)}]
		Sets the homogeneous solutions (together with periodic functions) for non-block master integrals.
		\texttt{sols} is a list of rules of form \texttt{m $\to$ homo}, where \texttt{m} is a master integral and \texttt{homo} is an expression depending on $\nu$.
\end{description}
For a more detailed description of the class \TMasters, the reader is referred to the \DREAM\ documentation.

\subsection{Other useful functions}

Additionally, \DREAM\ contains several useful functions.
The function \texttt{HomogeneousSolution} builds a homogeneous solution for a non-block master integral $J(\nu)$ by its \emph{certificate} $C(\nu)$:
\begin{equation}
	J(\nu+1) = C(\nu) \, J(\nu) + R(\nu).
\end{equation}
The \texttt{HomogeneousSolution} call looks like:
\begin{mmaCell}[moredefined={HomogeneousSolution}]{Input}
	HomogeneousSolution\pbl\mmaFrac{1}{\mmaSup{\mmaUnd{\(\nu\)}}{4}}, \mmaUnd{\(\nu\)}\pbr
\end{mmaCell}
\begin{mmaCell}{Output}
	\mmaFrac{1}{\mmaSup{Gamma[\(\nu\)]}{4}}
\end{mmaCell}

\texttt{FindAsymptotics} returns asymptotics of the function $f(\nu)$ by its certificate $c(\nu) = f(\nu+1) / f(\nu)$.
For example, for $f(\nu) = \Gamma(\nu) / \Gamma(\nu+1)$:
\begin{mmaCell}[moredefined={FindAsymptotics}]{Input}
	FindAsymptotics\pbl\mmaFrac{\mmaUnd{\(\nu\)}}{\mmaUnd{\(\nu\)}\,+\,1}, \mmaUnd{\(\nu\)}\,\(\to\infty\)\pbr
\end{mmaCell}
\begin{mmaCell}{Output}
	\{1, 1, 0\}
\end{mmaCell}
\texttt{FindAsymptotics} also implements the already mentioned Tulyakov's algorithm, Ref.~\cite{Tulyakov:2011}, and can find asymptotics of the slowest solution of the recurrence relation:
\begin{equation}
	\sum_{i=0}^{N} a_i(\nu) f(\nu+i) + R(\nu) = 0,
\end{equation}
given its coefficients $\{a_N, a_{N-1}, \ldots, a_0, R\}$.
For example:
\begin{mmaCell}[moredefined={FindAsymptotics}]{Input}
	FindAsymptotics[\{-45\,(4\mmaUnd{\(\nu\)}\,+\,1)(4\mmaUnd{\(\nu\)}\,+\,3), 2(549\mmaSup{\mmaUnd{\(\nu\)}}{2}\,+\,16), 400(3\mmaUnd{\(\nu\)}\,-\,2)(3\mmaUnd{\(\nu\)}\,-\,1), 0\}, \mmaUnd{\(\nu\)}\,\(\to\infty\)]
\end{mmaCell}
\begin{mmaCell}{Output}
	\big\{\mmaFrac{25}{8}, 1, 0\big\}
\end{mmaCell}

The function \texttt{ConvergenceAccelerate} implements the convergence acceleration algorithm which has been described in Ref.~\cite{LeeMingulov:2016:SummerTime}.
It returns an estimation of the sequence limit given its terms and its asymptotic behavior in terms of the aforementioned parameters $\{q, \alpha\}$.
For instance:
\begin{mmaCell}[moredefined={ConvergenceAccelerate,Accumulate}, morelocal={n}]{Input}
	l\,=\,N[Table[\mmaSup{n}{-2}, \{n, 1, 50\}], 50];
	l\,=\,Accumulate[l];
	r\,=\,ConvergenceAccelerate[l, All, \{1, 2\}]
\end{mmaCell}
\begin{mmaCell}{Output}
	1.6449340668482264364724151666460317614055416321753
\end{mmaCell}
\begin{mmaCell}[moredefined={r}]{Input}
	r\,-\,Zeta[2]
\end{mmaCell}
\begin{mmaCell}{Output}
	6.5721865917309685\(\times\)\mmaSup{10}{-33}
\end{mmaCell}

\section{Conclusion}

In this paper we have presented the \WMathematica\ package \DREAM\ dedicated to the application of the DRA method to the master integrals topologies containing the sectors with several master integrals.
This package automates the construction of the inhomogeneous parts of the solutions in terms of nested sums.
The package can compute master integrals with arbitrarily high precision (up to several thousands digits) and therefore can be used in conjunction with the \texttt{PSLQ} algorithm in order to recover answers in terms of conventional transcendental constants.
\DREAM\ can also automatically find the homogeneous solutions of the first-order recurrence relations in terms of $\Gamma$-functions and use the custom procedures for the computation of the homogeneous solutions of the higher-order recurrence relations defined by the user.
A method of building of these solutions has been proposed in the parallel paper \cite{LeeMingulov:Meromorphic}.

We provide two nontrivial examples of the package applications.
First, we calculate the master integrals entering the width of parapositronium decay into four photons. Using the \texttt{PSLQ} algorithm, we obtain for the first time the analytical form for this width.
Then we apply the package to computation of the fully massive four-loop tadpoles of cat-eye topology, \cref{eq:tadpole-topo}.
On this example we demonstrate how the custom objects for the calculation of homogeneous solutions of higher-order recurrence relations can be defined and embedded into the \DREAM\ object hierarchy.
We obtain high-precision numerical results which agree perfectly with the known ones \cite{Laporta2002,SchroderVuorinen2003,Czakon:2005:4loop-tadpoles}.
The examples considered in the present paper may seem to be a bit artificial.
The reason is that the families of one-scale integrals which are naturally treated by the DRA method tend to contain nontrivial sectors with several master integrals at higher loop level than those of the multi-scale integrals.
Therefore, using this method at its full strength requires a powerful IBP-reduction at hand.
One of the problems that seems to be perfect for the application of the presented algorithm is the problem of calculation of the four-loop massless form-factor master integrals \cite{HennSmirnovSteinhauser:4loop-form-factors,HennLeeSmirnovSteinhauser:4loop-form-factors,LeeSmirnovSteinhauser:4loop-form-factors-nf2,ManteuffelSchabinger:4loop-form-factors-nf3}.
In principle, the new package is applicable to the general problem of finding solutions of arbitrary-order recurrence relations with coefficients being rational functions of $\nu$,
the problem which can be of interest not only in analytical multiloop calculations, but also in some fields of mathematics.

\section*{Acknowledgments}

The work of the authors was supported in part by the RFBR grant \#17-02-00830 and by the grant of the Basis Foundation for theoretical physics.

\appendix

\section{Computed values of four-loop massive tadpoles}

In this appendix we present the results for master integrals $T_{4,5,9,10}$.
Analytical results were obtained with help of some heuristic conjectures and \texttt{PSLQ} algorithm \cite{FergusonBailey:1992:PSLQ}.

The master integrals at $\D=3-2\epsilon$:
\begin{align}
	\frac{T_4(3/2-\epsilon)}{T_1(3/2-\epsilon)} = &\ 
	-\frac{45}{16\epsilon}+\left[\frac{25}{2}\log\frac{5}{2}-\frac{45}{4}\right]
	-\left[\frac{45}{2}\text{Li}_2\left(\frac{3}{5}\right)+\frac{5\pi^2}{4}+30 \log ^2\frac{5}{2}-50 \log \frac{5}{2}+\frac{315}{4}\right]\epsilon \nonumber\\
	&\ + \bigg[5\,\text{Li}_3\!\left(\frac{1}{5}\right)+125\,\text{Li}_3\!\left(\frac{2}{5}\right)+\frac{135}{2}\text{Li}_3\!\left(\frac{3}{5}\right)-5\, \text{Li}_3\!\left(-\frac{1}{5}\right)+\frac{5}{2}\text{Li}_3\!\left(-\frac{4}{5}\right) \nonumber\\
	&\ \ \ \ \ \ +10\,\text{Li}_2\!\left(\frac{1}{5}\right)\log \frac{5}{2}-5\,\text{Li}_2\!\left(\frac{3}{5}\right) \log \frac{5}{2}-10\, \text{Li}_2\!\left(-\frac{1}{5}\right) \log\frac{5}{2}+\frac{5 \zeta_3}{4}+\frac{170}{3} \log ^3\frac{5}{2} \nonumber\\
	&\ \ \ \ \ \ -\frac{5}{12} \log ^3\frac{5}{4}-\frac{145}{2} \log \frac{5}{3} \log ^2\frac{5}{2}+\frac{155}{6} \pi ^2 \log\frac{5}{2}-\frac{5}{12}\pi^2 \log\frac{5}{4}-90\,\text{Li}_2\!\left(\frac{3}{5}\right)-5 \pi ^2 \nonumber\\
	&\ \ \ \ \ \ -120 \log ^2\frac{5}{2}+350 \log \frac{5}{2}-495 \bigg] \epsilon^2 + \mathcal{O}(\epsilon^3).
	\\
	\frac{T_5(3/2-\epsilon)}{T_1(3/2-\epsilon)} =&\ \frac{1}{8\epsilon} + \left[ \frac{3}{8}-\frac{1}{2} \log \frac{5}{2} \right] + \left[ \frac{1}{2}\text{Li}_2\!\left(\frac{3}{5}\right)+\frac{\pi^2}{12}+\log ^2\frac{5}{2}-2 \log \frac{5}{2} +\frac{3}{2} \right]\epsilon \nonumber\\
	&\ + \bigg[ \text{Li}_3\!\left(\frac{1}{5}\right)-5\,\text{Li}_3\!\left(\frac{2}{5}\right)-\frac{3}{2} \text{Li}_3\!\left(\frac{3}{5}\right)-\text{Li}_3\!\left(-\frac{1}{5}\right)+\frac{1}{2}\text{Li}_3\!\left(-\frac{4}{5}\right)+2\, \text{Li}_2\!\left(\frac{1}{5}\right) \log \frac{5}{2} \nonumber\\
	&\ \ \ \ \ \ -\text{Li}_2\!\left(\frac{3}{5}\right) \log \frac{5}{2}-2\,\text{Li}_2\!\left(-\frac{1}{5}\right) \log \frac{5}{2}-\frac{7 \zeta_3}{4}-\frac{2}{3} \log^3\frac{5}{2}-\frac{1}{12} \log ^3\frac{5}{4}+\frac{1}{2} \log \frac{5}{3} \log ^2\frac{5}{2} \nonumber\\
	&\ \ \ \ \ \ -\frac{5}{6} \pi ^2 \log \frac{5}{2}-\frac{1}{12} \pi ^2 \log \frac{5}{4}+ 6\,\text{Li}_2\!\left(\frac{3}{5}\right)+6 \log ^2\frac{5}{2}-6 \log \frac{5}{2} +\frac{25}{2}\bigg]\epsilon^2  + \mathcal{O}(\epsilon^3). \\
	\frac{T_9(3/2-\epsilon)}{T_1(3/2-\epsilon)} =&\ -0.07753725611432577874148759575094751360307470426691149891271881184\ldots \nonumber\\
	&\ +0.7850657447681336205546345615905118994345599157639593394166547118\ldots\epsilon \nonumber\\
	&\ -3.87222987092117894170216184691888444953509643858272522415\ldots\epsilon^2  + \mathcal{O}(\epsilon^3). \\
	\frac{T_{10}(3/2-\epsilon)}{T_1(3/2-\epsilon)} =&\ 0.007002151731638775971068172338414983664089137661364157458988760063656\ldots \nonumber\\
	&\ -0.034856950209820317451958160950917930710715091267297184432\ldots\epsilon  + \mathcal{O}(\epsilon^2).
\end{align}
The master integrals at $\D=4-2\epsilon$:
\begin{align}
	\frac{T_4(2-\epsilon)}{T_1(2-\epsilon)} =&\ \frac{5}{2} + \frac{5}{3}\epsilon + \frac{5}{144}\epsilon^2 + \frac{625}{864}\epsilon^3 +
	16.146469213466433695263636425983589657\ldots\epsilon^4 + \mathcal{O}(\epsilon^5). \\
	\frac{T_5(2-\epsilon)}{T_1(2-\epsilon)} =&\ -\frac{7}{6}\epsilon + \frac{3}{16}\epsilon^2 + \frac{83}{96}\epsilon^3 + 
	15.372633295802505967112028785600581887624\ldots\epsilon^4 + \mathcal{O}(\epsilon^5). \\
	\frac{T_9(2-\epsilon)}{T_1(2-\epsilon)} =&\ 
	\frac14+\frac43\epsilon + \left[\zeta _3-\frac{36}{5}\text{SR2a}+\frac{2 \pi ^2}{5}+\frac{25}{4}\right]\epsilon^2 + \bigg[\frac{\pi^4}{60} +\frac{4 \pi ^3}{15 \sqrt{3}} - \frac{107}{33}\zeta_3 -\frac{1}{5} 2 \pi ^2 \log 3 \nonumber\\
	&\ \ \ \ \ -\frac{144}{11}\text{SR3b} + \frac{72}{55}\pi\sqrt{3}\,\text{SR2a}+\frac{216}{55} \text{SR2a} \log 3 -\frac{144}{5}\text{SR2a} +\frac{8 \pi ^2}{5}+\frac{161}{6} \bigg] \epsilon^3  + \mathcal{O}(\epsilon^4).
	\\
	\frac{T_{10}(2-\epsilon)}{T_1(2-\epsilon)} =&\ \frac{\zeta_3}{2} \epsilon^2 + \left[ \frac{3}{4}\zeta_4 - \frac{\zeta_3}{2}\right] \epsilon^3  + \mathcal{O}(\epsilon^4).
\end{align}
Here we used the following constants:
\begin{align}
	\text{SR2a} &= \sum_{n=0}^{\infty} \frac{1}{(6 n+1)^2} = \frac{1}{36} \psi^{(1)}\!\left(\frac{1}{6}\right) = 1.03662536367637941436613053239706611913452559\ldots, \\
	\text{SR3b} &= \sum_{n=0}^{\infty} \sum_{k=n}^{\infty} \frac{1}{(6 n+1)^2} \left( \frac{1}{6 k+1}-\frac{1}{6 k+3} \right) = 0.729646011548050978211802034582339615\ldots
\end{align}
The coefficients of expansions of $T_{4,9}$ at $\D=4-2\epsilon$ and $3-2\epsilon$ have already been numerically found in Refs.~\cite{Laporta2002,SchroderVuorinen2003}.
The analytical of form the coefficients at $\D=4-2\epsilon$ up to the $3$-rd order by $\epsilon$ have been published in Ref.~\cite{Czakon:2005:4loop-tadpoles}.
Our results are in the perfect agreement with the already known ones.

\section*{References}


\end{document}